\documentclass[letterpaper,10pt,final,conference]{IEEEtran}
\usepackage[latin1]{inputenc}
\usepackage{graphicx}
\usepackage[english]{babel}
\usepackage{subfigure}
\usepackage{verbatim}
\usepackage{amsmath}
\usepackage{textcomp}
\usepackage{amssymb}
\usepackage{amsfonts}
\usepackage[normalem]{ulem}
\usepackage{balance}

\newcommand{\NC}[2]{
\left(\!\!\begin{array}{c}#1\\#2\end{array}\!\!\right)
}

\providecommand{\LyX}{L\kern-.1667em\lower.25em\hbox{Y}\kern-.125emX\@}

\begin{document}

\title{High-Rate Short-Block LDPC Codes for Iterative Decoding with
  Applications to High-Density Magnetic Recording Channels} \author{
  \authorblockN{Damián A. Morero, Graciela Corral-Briones,
    Carmen Rodríguez, and Mario R. Hueda} \\
  \authorblockA { Digital Communications Research
    Laboratory - National University of Cordoba - CONICET\\
    Av. Vélez Sarsfield 1611 - Córdoba (X5016GCA) - Argentina\\
    Emails: dmorero, gcorral, cer, mhueda@com.uncor.edu} %
  \thanks{This work was supported in part by SeCyT - UNC} }

\maketitle
 
\begin{abstract}
  This paper investigates the \textit{Triangle Single Parity Check}
  (T/SPC) code, a novel class of high-rate low-complexity LDPC codes.
  T/SPC is a regular, soft decodable, linear-time encodable/decodable
  code.  Compared to previous high-rate and low-complexity LDPC codes,
  such as the well-known \textit{Turbo Product Code / Single Parity
    Check} (TPC/SPC), T/SPC provides higher code rates, shorter code
  words, and lower complexity.  This makes T/SPC very attractive for
  practical implementation on integrated circuits.

  In addition, we analyze the performance of iterative decoders based
  on a soft-input soft-output (SISO) equalizer using T/SPC over
  high-density perpendicular magnetic recording channels. Computer
  simulations show that the proposed scheme is able to achieve a gain
  of up to 0.3~dB over TPC/SPC codes with a significant reduction of
  implementation complexity.  
\end{abstract}

\section{Introduction}
Iterative decoders based on soft-input soft-output equalizers with
powerful error correction codes (ECC) -such as low density parity
check (LDPC)- are considered in the literature to be suitable for
coping with intersymbol interference and noise in high-speed
transmission systems~\cite{STC2007,MSS2002,HK2007}. The potential of
this architecture is high, but it requires additional improvements in
order to be applicable to the next generation of magnetic recording
systems. There are mainly two reasons for this: the high complexity of
implementing high-speed iterative receivers in integrated circuits
(e.g., 4 Gb/s or higher), and the potential \textit{error floor}
problem of fully iterative decoding
solutions~\cite{SSKK,CH2006,SC2006}. The latter problem is exacerbated
by the difficulty of evaluating performance at very low bit error
rates (BERs).

One interesting alternative that offers a good tradeoff between
complexity and performance is the combination of \textit{(i)} an
iterative scheme based on a SISO equalizer with an \textit{inner}
high-rate low-complexity LDPC code and \textit{(ii)} a powerful
\textit{outer} code (such as RS, BCH, or Goppa), as shown in
Fig.~1~\cite{MOS2004}. In this scheme, the inner LDPC code should
achieve very high rates with low complexity as well as provide good
error statistics to the outer ECC. It has been shown that the
\textit{Turbo Product Code / Single Parity Check} (TPC/SPC) is a
suitable candidate to be used as inner code~\cite{LNKG2002,VK2005}.
The main features of the TPC/SPC codes are: minimum distance
$d_{H}=4$, no length-4 cycles, linear-time encodable and decodable
(i.e., low complexity implementation), and high code rate for
relatively short codewords \cite{LNKG2002,VK2005,LNG2004}. In magnetic
recording systems, where the sector size is fixed, the use of LDPC
with shorter code words provides benefits in an iterative
architecture. This is because the combination of interleaving with
\textit{various} code words per sector gives rise to interleaving
gain.  Therefore, the design of high rate short block LDPC codes holds
great interest for iterative receivers in high-density magnetic
recording systems.

In this paper we investigate the \textit{Triangle Single Parity Check}
(T/SPC) code, a novel LDPC code that can be derived from combinatorial
design criteria~\cite{VK2005,V2001,VKK2002,LK2002}. T/SPC is suitable
as inner code in the iterative decoding scheme shown in Fig. 1.
Compared to TPC/SPC, the proposed T/SPC exhibits:
\begin{itemize}
\item half code length for a given code rate;
\item half parity check nodes for a given code rate;
\item higher code rates for a given sector length;
\item lower minimum distance ($d_{H}=3$).
\end{itemize}

Note that T/SPC is able to provide a significant reduction of complexity but
this comes at the expense of a lower minimum distance when compared to
TPC/SPC.  However, as it will be shown later, combining a lower code length
together with an interleaver allows T/SPC to achieve a significant
interleaving gain. Furthermore, numerical results show that T/SPC is able to
provide a 0.3~dB gain over TPC/SPC in magnetic recording systems.

The rest of the paper is organized as follows. The system model is presented
in Section~\ref{sec:sysmodel}. T/SPC code is analyzed in
Section~\ref{sec:tspc}.  Section~\ref{sec:perfstim} evaluates the performance
of T/SPC and TPC/SPC. Finally, conclusions are drawn in
Section~\ref{sec:concl}.

\section{System Model} \label{sec:sysmodel}

\begin{figure}[t]
{\centering \resizebox*{1.00\columnwidth}{!}{\includegraphics{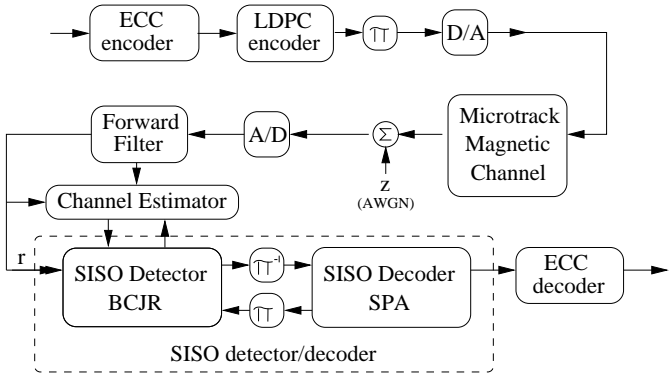}}\par}
\caption{\label{fig:SystemModel} System Model}
\end{figure}

The system model under study is shown in Fig.~\ref{fig:SystemModel}.
Information bits are first encoded with an ECC such as RS or BCH code (i.e.,
the outer code), and the output of the ECC is then encoded with an LDPC code
(i.e., the inner code). LDPC codes considered here are T/SPC and TPC/SPC.
Codewords at the LDPC output are interleaved using a random-design
interleaver. At the receiver side, samples are first processed by an adaptive
linear feed forward filter (FFF) in order to adapt the channel response to the
desired target response. Finally, the FFF output is used by the iterative
detector to estimate the transmitted bit sequence. The SISO detector is
implemented using the Max-Log-MAP approximation of the BCJR
algorithm~\cite{RHV1997}. The SISO decoder runs with two iterations of the
min-sum approximation of Sum-Product-Algorithm (SPA)~\cite{LNG2004}.

\subsection{Channel Model}
The microtrack model~\cite{MCSW2004} is used for the magnetic
recording channel with signal dependent transition noise. The received
signal is given by
\begin{eqnarray} 
  s(t) = \frac{1}{N_t} \sum_k \sum_{n=1}^{N_t} b_k \cdot h(t-T_k
  -\tau_{k,n}) + z(t) 
\end{eqnarray}
where:
\begin{itemize}
\item $h(t)$ is the transition response of a simple microtrack.  $h(t) =
  V\cdot erf\left( \frac{2\sqrt{ln(2)}\cdot t}{PW_{50}} \right)$, where $V$ is
  the transition amplitude, $PW_{50}$ is defined as the width of the
  derivative of $h(t)$ at half its peak amplitude.
\item $b_k \in\{-1, +1\}$ represents the transitions direction.
\item $T_k$ is the ideal time in which the $k$th transition takes place.
\item $\tau_{k,n}$ is the random shift of the $k$th transition on the $n$th
  microtrack. 
\item $N_t$ is the total number of microtracks ($N_t=2$ is used in this
  paper).
\item $z(t)$ is electronic noise.
\end{itemize}

The density is defined as $D=\frac{PW_{50}}{T}$, where $T$ is the bit period.
The transition jitter noise $\tau_{k,n}$ is modeled as an i.i.d. white
Gaussian random variable. The signal-to-noise ratio (SNR) is defined as $SNR =
\frac{1}{\sigma_z^2+\sigma_j^2}$ ($V=1$ is assumed), where $\sigma_z^2$ is the
power of the electronic noise samples, and $\sigma_j^2$ the power of the media
noise component~\cite{VK2005}. Media noise is defined by
$s_j(t)=s_{sj}(t)-s_s(t)$, where $s_{sj}(t)$ is the microtrack channel output
and $s_s(t)$ the jitter-free microtrack channel output.

\section{The Triangle/SPC Code} \label{sec:tspc}

\begin{figure}[t]
{\centering \resizebox*{0.90\columnwidth}{!}{\includegraphics{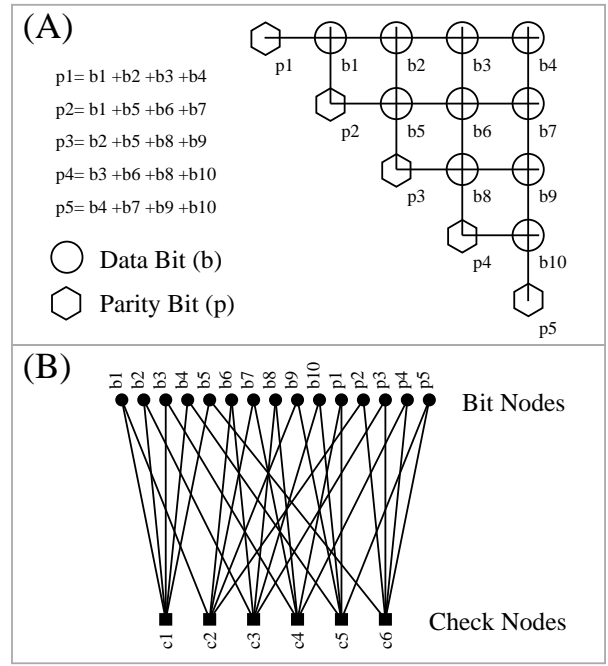}}\par}
\caption{\label{fig:TSPC00} A) Parity bits structure of a T/SPC(5) code B) Tanner graph of a 
T/SPC(5) code.}
\end{figure}

T/SPC code is a subclass of LDPC codes, obtained from the combination of
single parity check codes. A special case is the two-dimensional T/SPC,
denoted as $2$D-T/SPC, where the data and parity bits of the codeword are
arranged in a 2-dimensional triangular array as shown in
Fig.~\ref{fig:TSPC00}-A. Its Tanner graph is depicted in
Fig.~\ref{fig:TSPC00}-B.  Note that the 2D-T/SPC is a systematic code with the
single parity bits as redundancy. This code can be completely characterized by
the number of nodes in the edges of the triangle,
$N$. The parameters of a $2$D-T/SPC($N$) code are:\\[1mm]
\begin{tabular}{l l l}
  Minimum dist.:    &$d_{2,N}$& $= 3$            \\
  Code length:      &$n_{2,N}$& $= N(N+1)/2$     \\
  Code dimension:   &$k_{2,N}$& $= N(N-1)/2$     \\
  Code rate:        &$R_{2,N}$& $= (N-1)/(N+1)$  \\
  Check equations:  &$c_{2,N}$& $= N+1$          
\end{tabular}\\[1mm]

A $2$D-T/SPC code can be extended to an $M$-dimensional T/SPC ($M$D-T/SPC)
code by building an \textit{$M$-dimensional triangular array}.  The code
parameters of the resulting $M$D-T/SPC($N$) are:\\[1mm]
\begin{tabular}{l l l}
  Minimum dist.:   &$d_{M,N}$& $= M+1                  $ \\
  Code length:     &$n_{M,N}$& $= \sum_{i=1}^N n_{M-1,i} $ \\
  Code dimension:  &$k_{M,N}$& $= \sum_{i=1}^N k_{M-1,i} $ \\
  Code rate:       &$R_{M,N}$& $= k_{M,N}/n_{M,N}       $  \\
  Check equations: &$c_{M,N}$& $= \sum_{i=1}^N c_{M-1,i} $ \\
\end{tabular}\\[1mm]

We focus on the 2D-T/SPC code that achieves a higher code rate to code-length
ratio than the 2D-TPC/SPC. We first show that the 2D-T/SPC code, as the
2D-TPC/SPC, is a special case of \textit{Combinatorial Design Codes}, and
later the lower complexity of the T/SPC compared with the TPC/SPC.

\subsection{T/SPC and Combinatorial Design Codes }

The T/SPC can be analyzed using \textit{combinatorial design}. A combinatorial
design is an arrangement of a set of $m$ points into $n$ subsets, called
blocks, which satisfy certain regularity
constraints~\cite{VK2005,V2001,VKK2002,LK2002}. The \textit{covalency}
$\lambda_{v_1,v_2}$ of two points $v_1$ and $v_2$ is the mumber of blocks that
contain both of them. If $\lambda_{v_1,v_2}$ is the same for all pairs of
points, the design is said to be \textit{balanced}.  The number of points
contained in each block and the number of blocks each point is incident with,
are denoted by $\gamma$ and $\rho$, respectively. If $\gamma$ and $\rho$ are
the same for each block and point, respectively, the design is said to be
\textit{regular}.  A regular and balanced design is denoted as a
$(m,n,\rho,\gamma,\lambda)$-design. The \textit{incidence} matrix $M$ of the
combinatorial design has dimension $n\times m$, where $M_{i,j}=1$ if the point
$v_j$ is incident with the block $B_i$ and $M_{i,j}=0$ otherwise.

The transpose of the incidence matrix may be used as the parity check matrix
$H$ of an LDPC code. In this construction of an LDPC code, a point in a
combinatorial design corresponds to a \textit{check} node or a row in $H$, and
a block corresponds to a \textit{bit} node or a column in $H$. The row and
column weights of $H$ are $\rho$ and $\gamma$ respectively. A covalency
$\lambda<2$ guarantees the absence of length-4 cycles in the Tanner graph. For
example, the 2D-TPC/SPC(N,N-1)$^2$ can be constructed from a
$(2\rho,\rho^2,\rho,2,\{0,1\})$-design~\cite{LK2002} where $\rho=N$.

One of the most interesting classes of combinatorial design are the
\textit{Steiner systems}~\cite{V2001,MD1999}. A $(m,\gamma,t)$-Steiner-system
is a set $\mathcal{M}$ of m points, and a collection $\mathcal{N}$ of subsets
of $\mathcal{M}$ of size $\gamma$, called blocks, such that any subset of $t$
points of $\mathcal{M}$ is in exactly one of the blocks. The size of a Steiner
system, $n$, is defined as the number of blocks:
\begin{eqnarray}
  n=|\mathcal{N}| = \NC{m}{t} \Big/ \NC{\gamma}{t}
\end{eqnarray}

In \cite{V2001,VKK2002,MD1999} the construction and performance of LDPC codes
based on a $(m,3,t)$-Steiner-system (called Steiner triple systems), and a
particular case called Kirkman system, are analyzed.  We will focus on the
$(N-1,2,2)$-Steiner-system which produces a $(m,n,\rho,\gamma,\lambda)$-design
with parameters: $n=N(N+1)/2$, $m=N+1$, $\rho=N$, $\gamma=2$, $\lambda=1$. The
code obtained from the incidence matrix of this design is the 2D-T/SPC($N$).

\subsection{Complexity of T/SPC and TPC/SPC}

The code rate of a 2D-TPC/SPC($N$,$N-1$)$^2$ is given by
{\small \begin{eqnarray} 
  R_{TPC/SPC}&=& \left(\frac{N-1}{N}\right)^2  \\
  &=& \frac{N-1}{N+1} - \frac{1}{N(N+1)} + \frac{1}{N^2(N+1)} \nonumber \\
  &\approx& \frac{N-1}{N+1} = R_{T/SCP} \;\;\;\; for \;\; N>>1 \nonumber 
\end{eqnarray}}
Therefore, a simple complexity comparison between 2D-T/SPC and 2D-TPC/SPC can
be done considering the same parameter $N$ for both codes.

Given that complexity is mainly on the decoder, we will focus on the SPA-based
decoder. T/SPC requires approximately the same bit-nodes and check-nodes
computations as TPC/SPC. In general, the memory requirement and the
interconnection complexity are proportional to the number of edges ($N_e$) of
the code factor graph. Considering that the number of edges is $N_e=N(N+1)$
for T/SPC and $N_e=2N^2$ for TPC/SPC, we conclude that T/SPC is able to
provide a significant reduction of memory and interconnection complexity
($\sim 1/2$ for $N>>1$).

\section{Numerical Results} \label{sec:perfstim}

The performance of T/SPC and TPC/SPC is analyzed by using computer simulations
of the system described in Section~\ref{sec:sysmodel}.  Code rates of 0.94 and
0.969 are analyzed. We consider a perpendicular channel of density $D=3$ and
80\%~/~20\% jitter/electronic noise power (i.e.,
$\sigma_j^2/\sigma_z^2=0.8/0.2$).  A generalized partial response target with
4 taps (GPR4) is utilized.  The LDPC decoder performs two SPA iterations.
Random-design interleavers are used.

\begin{figure}[t]
{\centering \resizebox*{1.0\columnwidth}{!}{\includegraphics{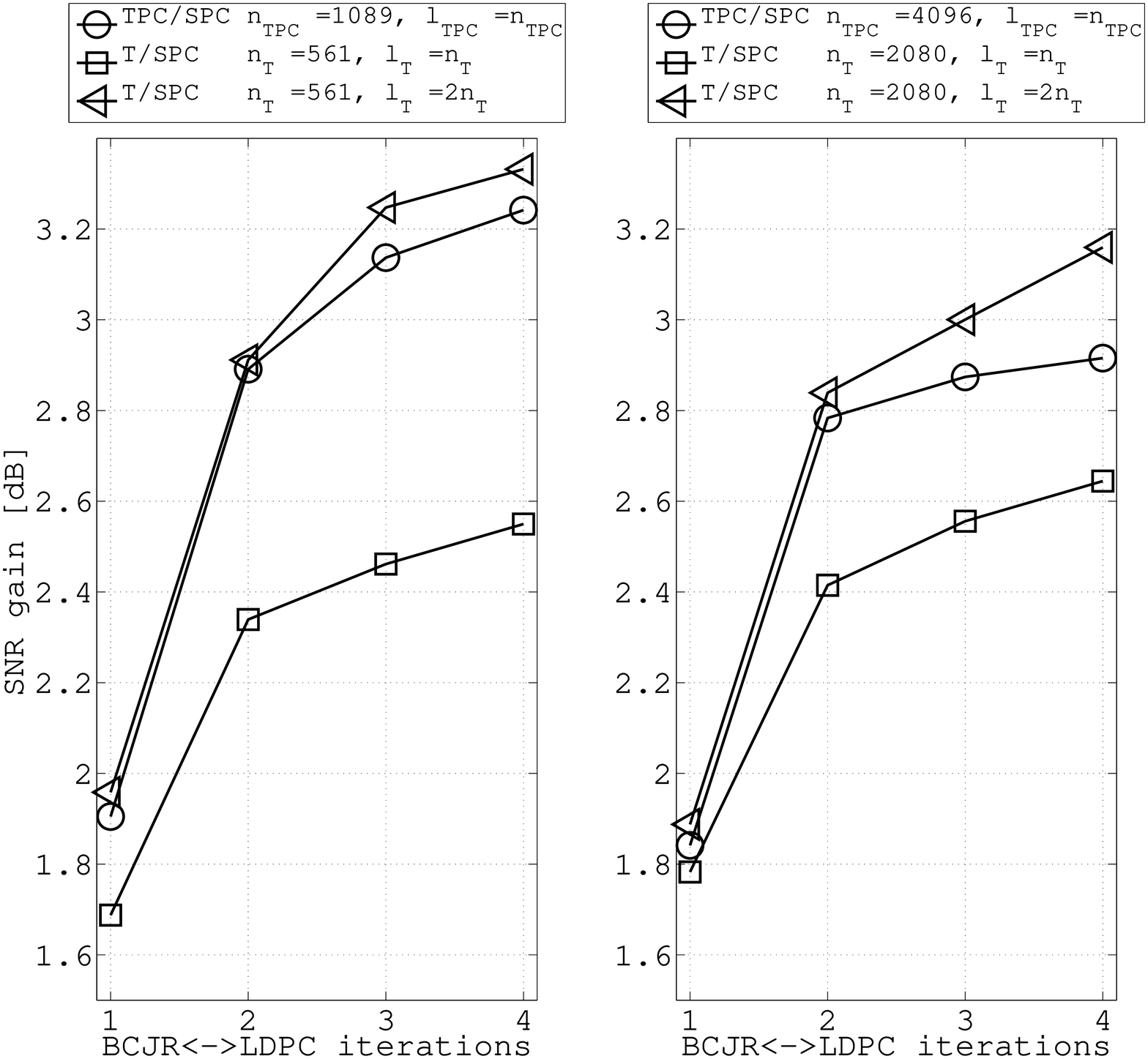}}\par}
\caption{\label{fig:SNRGAIN12} SNR gain vs. number of BCJR$\leftrightarrow$LDPC
  iterations at BER$=10^{-4}$. Code rates 0.94 (left) and 0.969 (right).
  \vspace{4mm}}
\end{figure}

\begin{figure}[t]
  {\centering \resizebox*{1.0\columnwidth}{!}{\includegraphics{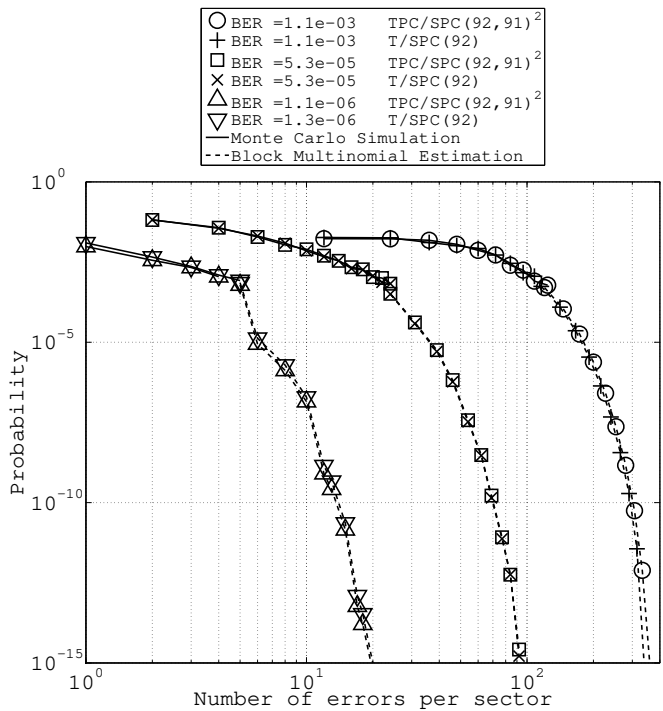}}\par}
  \caption{\label{fig:HIST1} Probability of the number of errors per
    sector over a PR[1~3~3~1] with 100$\%$ electronic noise. Four
    TPC/SPC($92,91$)$^2$ codes and eight T/SPC($92$) codes per sector.
    \vspace{4mm}
  }
\end{figure}
 
Let $l_T$ and $l_{TPC}$ be the interleaver lengths used on the T/SPC and
TPC/SPC codes respectively. Also, let $n_T$ and $n_{TPC}$ be the code-word
lengths for the T/SPC and TPC/SPC codes respectively. For TPC/SPC we use
$l_{TPC}=n_{TPC}$. On the other hand, we analyze T/SPC with two interleaver
lengths: $l_T=n_T$ and $l_T=2n_T\approx l_{TPC}$.  Figure~\ref{fig:SNRGAIN12}
shows the SNR gain vs.  number of BCJR$\leftrightarrow$LDPC iterations at
BER$=10^{-4}$.  The performance achieved by an uncoded BCJR equalizer is taken
as reference.  The resulting performance of T/SPC for $l_T=n_T$ is worse than
the one of TPC/SPC, mainly because of the T/SPC lower minimum distance.
However, for $l_T = 2n_T\approx l_{TPC}$ the interleaver gain compensates the
degradation caused by its lower minimum distance.  Furthermore, in some cases
note that T/SPC is able to provide a gain of around 0.27~dB over TPC/SPC owing
to the interleaving gain. Note that this gain is achieved with lower
complexity.

In the iterative-based receiver depicted in Fig.
\ref{fig:SystemModel}, the statistics of the bit errors at the output
of the inner code strongly affect the performance of the outer code.
Next we investigate the distribution of bit errors at the output of
the T/SPC and TPC/SPC with code rate 0.978 and two
BCJR$\leftrightarrow$LDPC iterations.  Sector size is 4KB and the
interleaving length is 1KB.  Figure~\ref{fig:HIST1} shows the
probability of the number of errors per sector for a partial response
channel PR[1~3~3~1] with 100$\%$ electronic Gaussian noise.  Solid
lines correspond to values derived from simulations, while dashed
lines represent estimates calculated by using the block-multinomial
model~\cite{KKHB2004}.  From Fig.~\ref{fig:HIST1} we note that, at a
given BER, the statistics of bit errors at the output of T/SPC are
practically the same as those observed with TPC/SPC.  Therefore, we
infer that the behavior of T/SPC and TPC/SPC in an iterative-based
receiver as shown in Fig. \ref{fig:SystemModel}, should be similar.
This topic, and other related issues, will be addressed in a future
work.

\balance
\section{Conclusions} \label{sec:concl}

We have proposed and investigated T/SPC, a novel high-rate LDPC code.
T/SPC provides higher code rates, shorter code words, and lower
complexity than TPC/SPC. Computer simulations of T/SPC on high-density
perpendicular magnetic recording channels have shown that T/SPC is
able to achieve a significant interleaving gain, outperforming TPC/SPC
by almost 0.3 dB. Our results suggest that the design of LDPC with high-rate,
low complexity and short code word is a promising research area for 
 next generation developments of magnetic recording devices.

\bibliographystyle{IEEEtran}
\bibliography{paper}

\end{document}